\documentclass[acmtog]{acmart}
\acmSubmissionID{2478}
\usepackage{booktabs} 
\usepackage[ruled]{algorithm2e} 

\SetAlFnt{\small}
\SetAlCapFnt{\small}
\SetAlCapNameFnt{\small}
\SetAlCapHSkip{0pt}
\usepackage{xcolor}
\usepackage{pifont}
\newcommand{\tablestyle}[2]{\setlength{\tabcolsep}{#1}\renewcommand{\arraystretch}{#2}\centering}
\definecolor{gold}{HTML}{FAE37F}
\definecolor{silver}{HTML}{D7D7D7}
\definecolor{bronze}{HTML}{EDBA91}
\definecolor{half}{HTML}{FFC107}
\newcommand{\au}[1]{\setlength{\fboxsep}{1pt}\colorbox{gold}{#1}}
\newcommand{\ag}[1]{\setlength{\fboxsep}{1pt}\colorbox{silver}{#1}}

\newcommand{\gcheck}[0]{\color{green}\ding{52}}
\newcommand{\rcross}[0]{\color{red}\ding{56}}
\newcommand{\hcheck}[0]{\color{half}\ding{52}}

\AtBeginDocument{%
  }

\copyrightyear{2026}
\acmYear{2026}
\setcopyright{cc}
\setcctype{by}
\acmConference[SA Conference Papers '26]{SIGGRAPH Asia 2026 Conference Papers}{December 01--04, 2026}{Kuala Lumpur, Malaysia}
\acmBooktitle{SIGGRAPH Asia 2026 Conference Papers (SA Conference Papers '26), December 01--04, 2026, Kuala Lumpur, Malaysia}
\acmDOI{10.1145/3829340.3842356}
\acmISBN{979-8-4007-2842-6/2026/12}

\newcommand{\proposed}{\textbf{Fallingwater}\xspace}

\begin{document}

\title{Personalizing Causal Audio-Driven Facial Motion via Dynamic Multi-modal Retrieval}

\author{Xuangeng Chu}
\affiliation{%
  \institution{The University of Tokyo}
  \city{Tokyo}
  \country{Japan}
}
\affiliation{%
  \institution{Codec Avatars Lab, Meta}
  \city{Pittsburgh}
  \country{United States of America}
}
\email{xuangeng.chu@mi.t.u-tokyo.ac.jp}

\author{Yu Han}
\affiliation{%
  \institution{Codec Avatars Lab, Meta}
  \city{Pittsburgh}
  \country{United States of America}
}
\email{vickyhan9999@gmail.com}

\author{Wei Mao}
\affiliation{%
  \institution{Codec Avatars Lab, Meta}
  \city{Pittsburgh}
  \country{United States of America}
}
\email{weimao@meta.com}

\author{Shih-En Wei}
\affiliation{%
  \institution{Sooth Labs}
  \city{Pittsburgh}
  \country{United States of America}
}
\email{weisteady@gmail.com}

\renewcommand{\shortauthors}{Chu et al.}

\begin{abstract}
Audio-driven facial animation is essential for immersive digital interaction, yet existing frameworks struggle to reconcile real-time streaming with high-fidelity personalization. Current methods either rely on latency-inducing audio look-ahead, or ask users to record scripted calibration sequences to pre-encode static identity embeddings that fail to capture dynamic idiosyncrasies.
We present an end-to-end framework for \textit{personalized} audio-driven facial motion generation, supporting causal, zero-lookahead streaming.
We introduce two key innovations: (1) a causal multi-resolution motion tokenizer that captures both global temporal context and high-frequency articulatory details, and (2) a multi-modal style retriever that extracts stylistic priors from \textit{unstructured} reference libraries by jointly querying ongoing audio and motion.
Unlike prior retrieval mechanisms restricted to curated, fixed-size, or audio-only style banks, our design accepts arbitrary footage of the target identity, enabling personalization from a handful of casually recorded clips.
By integrating these components, our method outperforms state-of-the-art approaches in lip-sync accuracy, identity consistency, and perceived realism, while preserving the streaming constraints of real-time telepresence.
Code is available at \url{https://github.com/xg-chu/Fallingwater}.
\end{abstract}



\begin{teaserfigure}
  \includegraphics[width=\textwidth]{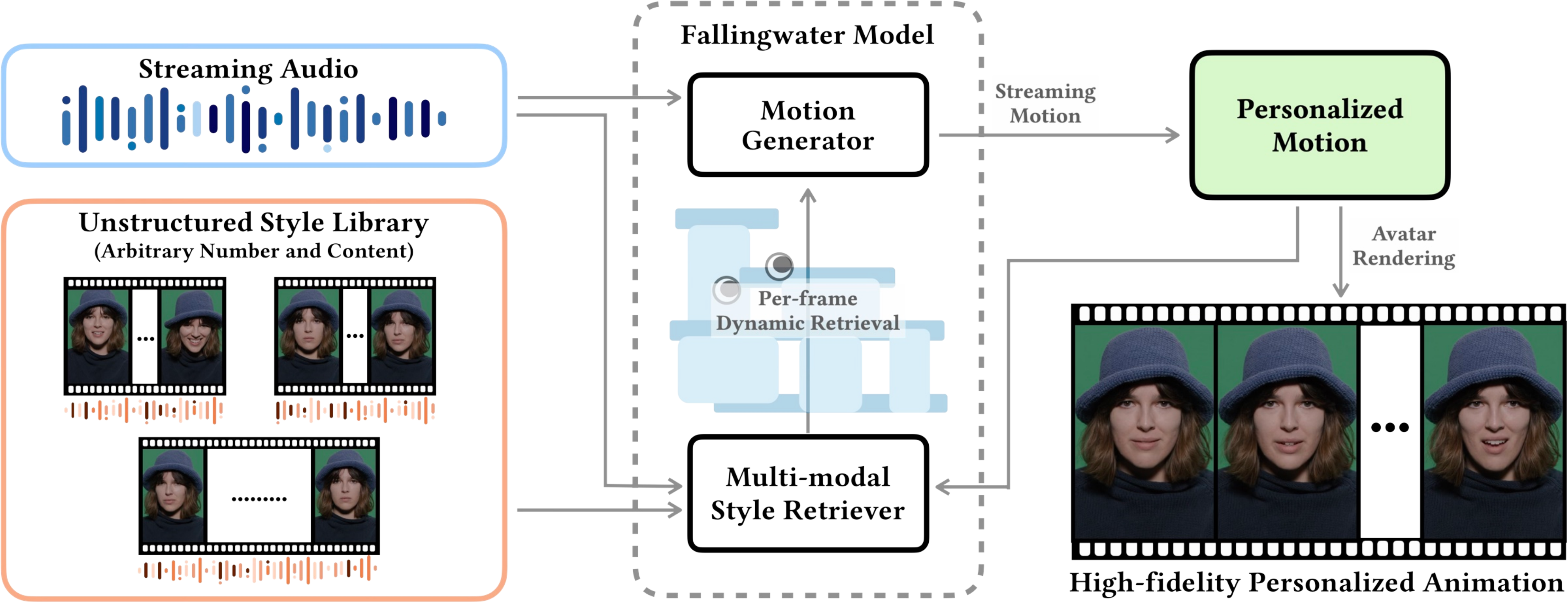}
  \caption{Given an audio stream and a small set of casually recorded clips of the target user (no scripted enrollment required), \proposed jointly queries audio and motion history to retrieve identity-specific style priors and generates personalized facial motion while supporting causal, zero-lookahead streaming at real-time throughput.}
  \Description{Pipeline diagram of Fallingwater: an audio stream and a small set of casually recorded reference clips are fed into a joint audio-motion retriever; the retrieved style prior conditions a causal autoregressive transformer that generates personalized facial motion frames one at a time.}
  \label{fig:teaser}
\end{teaserfigure}
\maketitle

\section{Introduction}
Live, identity-aware facial animation has become a core building block for immersive digital interaction, from telepresence in AR/VR \citep{thies2016face2face, lombardi2018deep, wei2019vr} to embodied virtual agents in games and broadcasting.
While camera-based motion capture offers a direct observation of facial dynamics, it is often impractical in deployment: head-mounted displays still face challenges from internal-camera occlusion and limited downstream bandwidth, and many everyday scenarios (mobile calls, low-bandwidth links, privacy-sensitive contexts) preclude continuous video capture altogether. The audio modality, in contrast, is immune to visual occlusion and far less bandwidth-intensive, while still carrying rich semantic and prosodic cues essential for driving facial animation \citep{karras2017audio}. Consequently, generating realistic, personalized motion from live audio streams has emerged as a critical capability for next-generation interactive systems.

Despite advancements in robust audio feature extraction \citep{defossez2024moshi}, the acoustic signal remains an under-determined driver of facial dynamics: it lacks the morphological cues needed to resolve idiosyncratic nuances such as asymmetric smirks, speaker-specific gingival exposure, or blinking patterns \citep{thambiraja2023imitator,kim2025memorytalker}. As a result, deterministic mappings suffer from ``regression to the mean'', producing over-smoothed motions that discard each subject's stylistic signature.

Existing methods for personalizing facial motion face three intertwined bottlenecks: latency, expressivity, and usability.
First, high-quality synthesis relies on either computationally intensive iterative sampling, as seen in diffusion-based models \citep{MotionStreamer_2025_ICCV} or audio look-ahead buffering \citep{fan2022faceformer, richard2021meshtalk, chu2024artalk}, both of which introduce systemic delays incompatible with real-time applications.
Second, the personalization signal is typically reduced to a fixed text descriptor \citep{lee2025audio} or a single pre-encoded motion snippet \citep{sun2024diffposetalk, chu2024artalk}: static, single-modality priors that fail to track style evolution across phonemes, prosody, or emotion, and generalize poorly to out-of-distribution styles.
Third, encoding such priors usually requires scripted calibration recordings, raising the practical cost of personalization.
Together, these expressivity and usability constraints prevent current architectures from scaling to \textit{unstructured} reference libraries: short, casually recorded clips of the target identity that already exist as everyday footage. Such footage encodes rich, jointly correlated audio-motion signatures that remain largely untapped by prior work.

To address these limitations, we propose a framework comprising two synergistic innovations. The first is a causal multi-resolution motion tokenizer that decomposes facial motion into a hierarchy of discrete latent codes across temporal resolutions, decoupling high-frequency articulation from lower-frequency expressive dynamics. While hierarchical tokenization has been explored before \citep{chu2024artalk}, prior decoders aggregate a full window of motion codes and are therefore non-causal, precluding streaming; we instead pair a strictly causal decoder with a causal upsampler that never references future codes, while the generator predicts motion tokens covering an upcoming time span from each arriving audio feature, producing frames immediately with no window-induced latency. The second is a multi-modal style retriever over \textit{unstructured} reference libraries: exploiting the recurrence of ID-specific stylistic patterns across utterances, we query the user's reference clips with both recent audio and motion history, unlike the audio-only or single-shot queries of prior style-bank methods. Being content-addressed and template-agnostic, the retriever accepts arbitrary references: a few short, casually recorded clips of the target identity suffice, with no scripted enrollment required. These two components are integrated into a single autoregressive transformer, which predicts motion tokens from audio, motion history, and retrieved style priors while supporting causal, zero-lookahead streaming.

Our main contributions are as follows:
\begin{itemize}
\item \textbf{A causal multi-resolution motion tokenizer} that enables zero-lookahead decoding and streaming synthesis.
\item \textbf{A multi-modal style retriever} that dynamically extracts ID priors from unstructured reference libraries.
\item A unified framework leveraging these innovations to achieve state-of-the-art expressive fidelity and identity consistency under causal, zero-lookahead streaming.
\end{itemize}

\section{Related Work}
\subsection{Audio-Driven Facial Motion Generation}
Audio-driven facial motion generation aims to synthesize realistic facial movements from acoustic signals. The field can be broadly categorized into direct mapping methods and latent-based frameworks.

\paragraph{Direct Mapping Methods} Early 2D portrait-video methods~\citep{suwajanakorn2017synthesizing, jamaludin2019you, prajwal2020lip, zhou2020makeittalk} struggle to enforce spatiotemporal consistency, motivating later works~\citep{ji2021audio, zhang2023sadtalker, ye2024real3d} that directly generate 3D facial representations. A common strategy predicts deformations on 3D mesh templates~\citep{cudeiro2019capture, fan2022faceformer, haque2023facexhubert, karras2017audio, nocentini2024scantalk, thambiraja2023imitator, xing2023codetalker}, whose fidelity is bounded by template expressiveness. To capture finer details, subsequent works~\citep{li2023efnerf, liu2022sspnerf, yao2022dfanerf, ye2023geneface, guo2021ad, shen2022dfrf, ye2023geneface++, li2025talkinggaussian, chen2024gstalker, aneja2024gaussianspeech, he2024emotalk3d} adopt Neural Radiance Fields (NeRF)~\citep{mildenhall2020nerf} or 3D Gaussian Splatting (3DGS)~\citep{kerbl20233dgs} for higher rendering quality; however, these volumetric or point-based representations are typically identity-specific, hindering cross-subject generalization.

\paragraph{Latent-Based Frameworks} Latent-based methods instead map audio to a compressed representation that is then decoded into 2D or 3D faces~\citep{aneja2024facetalk, sun2024diffposetalk, giebenhain2023nphm, qian2024gaussianavatars, ng2024audio2photoreal, richard2021meshtalk, lee2025audio}. TalkShow~\citep{yi2023talkshow} uses convolution-based regression to predict FLAME parameters~\citep{li2017learning}, while ARTalk~\citep{chu2024artalk} predicts hierarchical discrete latent codes autoregressively; the latter's decoder, however, operates non-causally over a full temporal window, precluding true streaming. Recent diffusion-based frameworks (e.g., FaceTalk~\citep{aneja2024facetalk}, DiffPoseTalk~\citep{sun2024diffposetalk}) achieve high-fidelity synthesis but rely on either full-sequence look-ahead or computationally intensive iterative sampling, making them unsuitable for causal, real-time use. Single-step diffusion~\citep{lee2025audio} reduces inference cost but still operates at chunk granularity, limiting true frame-by-frame streaming.

\paragraph{Our Approach.} In contrast, our multi-resolution motion tokenizer maps facial expressions to causal, hierarchical, discrete codes, enabling a transformer-based autoregressive model to predict motion while supporting causal, zero-lookahead streaming without sacrificing expressive fidelity.

\subsection{Personalization of Motion Generation}
To enhance controllability, existing frameworks~\citep{chu2024artalk,sun2024diffposetalk,lee2025audio,kim2025memorytalker,pan2025model,liu2025medtalk} rely on external style signals: identity-agnostic reference motions~\citep{chu2024artalk,sun2024diffposetalk}, categorical emotions~\citep{lee2025audio,liu2025medtalk}, or text descriptions~\citep{liu2025medtalk}. These signals are typically fixed across the generation process, failing to track stylistic evolution. MemoryTalker~\citep{kim2025memorytalker} addresses this with a learned style bank but retrieves entries using audio queries alone, leaving the motion modality untapped. Pan et al.~\citep{pan2025model} use key poses paired with a style network.

Despite these advances, existing methods still rely on structured or categorical inputs and struggle to preserve a subject's idiosyncratic ``personality.'' In contrast, our multi-modal style retriever adaptively extracts style priors from arbitrary, \textit{unstructured} facial videos of the target identity, that is, casually recorded clips with no pre-segmentation, labels, or fixed length.

\section{Methods}
\begin{figure*}[ht]
\begin{center}
\centerline{\includegraphics[width=1.0\linewidth]{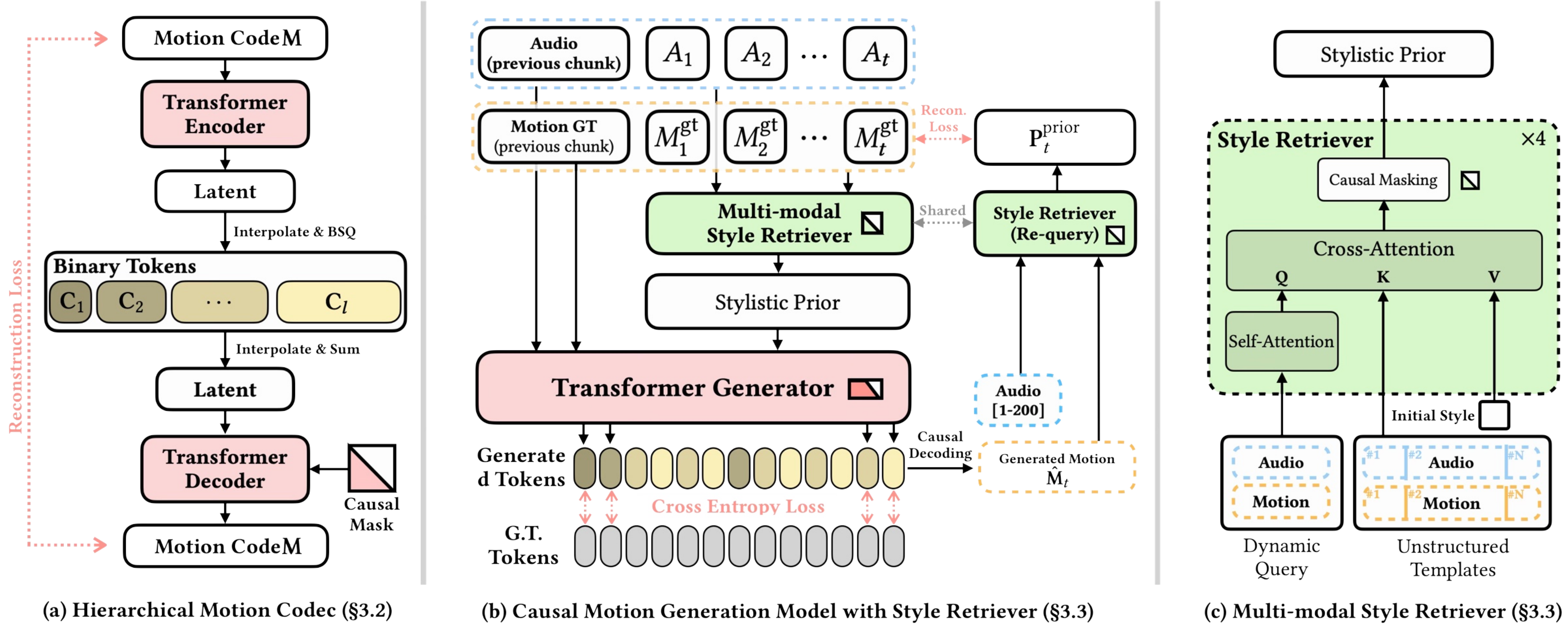}}
\caption{
\textbf{System overview of our method.}
Our framework consists of two primary components: (a) The \textbf{Hierarchical Motion Codec}, which compresses per-frame motion into multi-scale discrete tokens; and (b) the \textbf{Motion Generation with Style Retriever}, where an autoregressive transformer predicts these tokens in a streaming manner, conditioned on a multi-modal style retriever trained with a re-query strategy for generalization. (c) Detailed architecture of the style retriever, which operates on a set of unstructured templates.
}
\label{fig:overview}
\end{center}
\Description{System diagram with three labeled blocks: (a) the hierarchical motion codec compressing motion into multi-scale discrete tokens; (b) the streaming autoregressive generator conditioned on audio embeddings, previous-window motion history, and retrieved style priors; and (c) the multi-modal style retriever cross-attending over a library of unstructured templates.}
\end{figure*}

Our framework (Fig.~\ref{fig:overview}) enables personalized facial motion synthesis while supporting zero-lookahead streaming at real-time throughput.
We first define the motion space and acoustic features in \S\ref{sec:31}. \S\ref{sec:32} then details our causal hierarchical motion codec, which provides the discrete representational foundation for streaming. Finally, \S\ref{sec:33} describes the streaming autoregressive generator, its integration with the multi-modal style retriever, and the overall training scheme.

\subsection{Preliminaries}
\label{sec:31}

Our framework operates on any compact per-frame motion latent $M_i \in \mathbb{R}^{D}$, whether learned (e.g., the ``imitator'' latent space of~\citet{agrawal2025seamlessinteractiondyadicaudiovisual} or a VAE-based motion codec~\citep{lombardi2018deep}) or parametric (e.g., FLAME~\citep{li2017learning}). We instantiate $M_i$ with the imitator representation, concatenating facial expression $\mathbf{e}_i \in \mathbb{R}^{128}$, head rotation $\mathbf{r}_i^\text{head} \in \mathbb{R}^3$, head translation $\mathbf{t}_i^\text{head} \in \mathbb{R}^3$, and shoulder translation $\mathbf{t}_i^\text{shoulder} \in \mathbb{R}^3$ into $M_i \in \mathbb{R}^{137}$.

For audio, we extract embeddings using a pre-trained Mimi backbone~\citep{defossez2024moshi}, which encodes raw audio at $12.5$\,Hz into a $512$-dimensional latent stream. We then resample these latents by nearest-neighbor interpolation to the motion frame rate, yielding $A_i \in \mathbb{R}^{512}$ time-aligned with $M_i$.

\subsection{Causal Hierarchical Motion Codec}
\label{sec:32}

\begin{algorithm}[t]
\caption{\textbf{Hierarchical Motion Codec}}
\label{alg:hiera_codec}
\begin{center}
\begin{tabular}{ll}
\toprule
\multicolumn{2}{l}{\textbf{Temporal Hierarchical Encoding}} \\
\midrule
\textbf{Inputs:} Motion sequence $\mathbf{M}=[M_1,...,M_T]$, Hierarchy depth $L$, & \\
\quad \quad \quad \, code lengths $r_1 < \dots < r_L$ (coarse to fine). \\
\textbf{Output:} Motion tokens $\mathbf{C} = [\mathbf{C}_1, \dots, \mathbf{C}_L]$ & \\
$f = \text{EncoderTransformer}(\mathbf{M}) \in \mathbb{R}^{T\times d}$ \\
\textbf{for} $i = 1, \dots, L$: \\
\quad $\mathbf{C}_i = \text{BSQ}(\text{Interp}(f, r_i))$ \\
\quad $f \leftarrow f - \text{Interp}(\text{Dequantize}(\mathbf{C}_i),\, T)$ \\
\textbf{return} $\mathbf{C}$ \\
\midrule
\multicolumn{2}{l}{\textbf{Temporal Hierarchical Causal Decoding}} \\
\midrule
\multicolumn{2}{l}{\textbf{Inputs:} Motion tokens $\mathbf{C} = [\mathbf{C}_1, ..., \mathbf{C}_L]$} \\
\textbf{Output:} Motion sequence $\mathbf{M}=[M_1,...,M_T]$ & \\
$\hat{f} = 0$ & \\
\textbf{for} $i$ \textbf{in} $[1,...,L]$ \textbf{:} & \\
\quad \quad $\hat{z}_i = \text{CausalInterp}(\mathbf{C}_i, T)$ & \\
\quad \quad $\hat{f} = \hat{f} + \hat{z}_i$ & \\
$\mathbf{M} = \text{DecoderTransformer}(\hat{f}, \text{causal\_mask})$ & \\
\textbf{return} $\mathbf{M}$ &  \\
\bottomrule
\end{tabular}
\end{center}
\end{algorithm}

Inspired by the hierarchical principles of \citep{chu2024artalk, tian2024visualautoregressivemodelingscalable} that decouple fine-grained details from global dynamics, we present a multi-resolution temporal autoencoder that compresses each motion window into a hierarchy of discrete tokens at increasing temporal resolutions. Compared with prior hierarchical designs, our reformulated decoder enforces strict causality in both attention and upsampling, enabling true streaming synthesis without sacrificing reconstruction fidelity.

\paragraph{Temporal Hierarchical Encoder}
We partition the motion sequence into non-overlapping temporal windows of length $T$. For each window $\mathbf{M}=[M_1,\dots,M_T]$, we encode it into $p$-dimensional discrete codes $\mathbf{C} = [\mathbf{C}_1, \dots, \mathbf{C}_L]$ across $L$ levels, where each $\mathbf{C}_l \in \{0,1\}^{r_l \times p}$ corresponds to temporal resolution $r_l$.
We employ a residual quantization strategy to progressively minimize quantization error. Under this scheme, finer levels preserve high-frequency signals necessary for subtle motions, while the coarser levels utilize temporal downsampling to capture long-term prosody and expression trends.
To ensure a robust latent space, we apply Binary Spherical Quantization (BSQ) \citep{zhao2024imagevideotokenizationbinary}, which projects motion features onto a hypersphere and represents each token as an implicit binary codebook of size $2^p$, yielding stable training without codebook collapse and supporting an exponentially large code space at modest memory cost.
The full procedure is detailed in Algorithm \ref{alg:hiera_codec}.

\paragraph{Causal Decoder}
Traditional multi-scale encoder-decoders apply bidirectional attention during decoding, which relies on future motion frames and is therefore non-causal.
To enable streaming generation, we apply a triangular causal attention mask within the decoder, restricting it to tokens generated prior to the current frame and thereby enabling zero-lookahead decoding.
We further replace linear upsampling (which references future positions) with floor-indexed nearest-neighbor interpolation that reads only from past positions.
This causal constraint is enforced only on the decoder. Since the encoder is not involved in real-time inference, we leave it non-causal so the discrete codes can encapsulate long-term, bidirectional context from the input motion sequences. Empirical evaluations show that our causal codec matches the reconstruction quality of chunk-based methods while eliminating their window-induced latency.

\paragraph{Training Objectives}
To train the BSQ encoder-decoder, we employ a hybrid loss that balances frame-wise reconstruction, temporal smoothness, and codebook utilization. On the pose-only subset $R_i = [\mathbf{r}_i^\text{head}, \mathbf{t}_i^\text{head}, \mathbf{t}_i^\text{shoulder}] \in \mathbb{R}^{9}$, the velocity loss $\mathcal{L}_\text{vel} = \tfrac{1}{T-1}\sum_{i=1}^{T-1} \|(\hat R_{i+1} - \hat R_i) - (R^\text{gt}_{i+1} - R^\text{gt}_i)\|^2$ supervises predicted pose velocity against the ground truth, while the jitter regularizer $\mathcal{L}_\text{jit} = \tfrac{1}{T-2}\sum_{i=1}^{T-2} \|\hat R_{i+2} - 2\hat R_{i+1} + \hat R_i\|^2$ penalizes the raw magnitude of predicted pose acceleration (without GT subtraction) to suppress chattering. We additionally apply BSQ's entropy regularization $\mathcal{L}_\text{entropy}$ strictly following \citet{zhao2024imagevideotokenizationbinary} to prevent codebook collapse. The final training objective is:
\begin{equation}
\label{eq:codec_loss}
\mathcal{L}_\text{codec} = \frac{1}{T}\sum_{i=1}^{T} \left\| \hat{M}_i - M^{\text{gt}}_i \right\|_1 + \lambda_{v} \mathcal{L}_\text{vel} + \lambda_{j} \mathcal{L}_\text{jit} + \lambda_{e} \mathcal{L}_\text{entropy}.
\end{equation}

\subsection{Streaming Motion Generation Model with Multi-modal Style Retriever}
\label{sec:33}

\begin{figure}
\begin{center}
\centerline{\includegraphics[width=1.0\columnwidth]{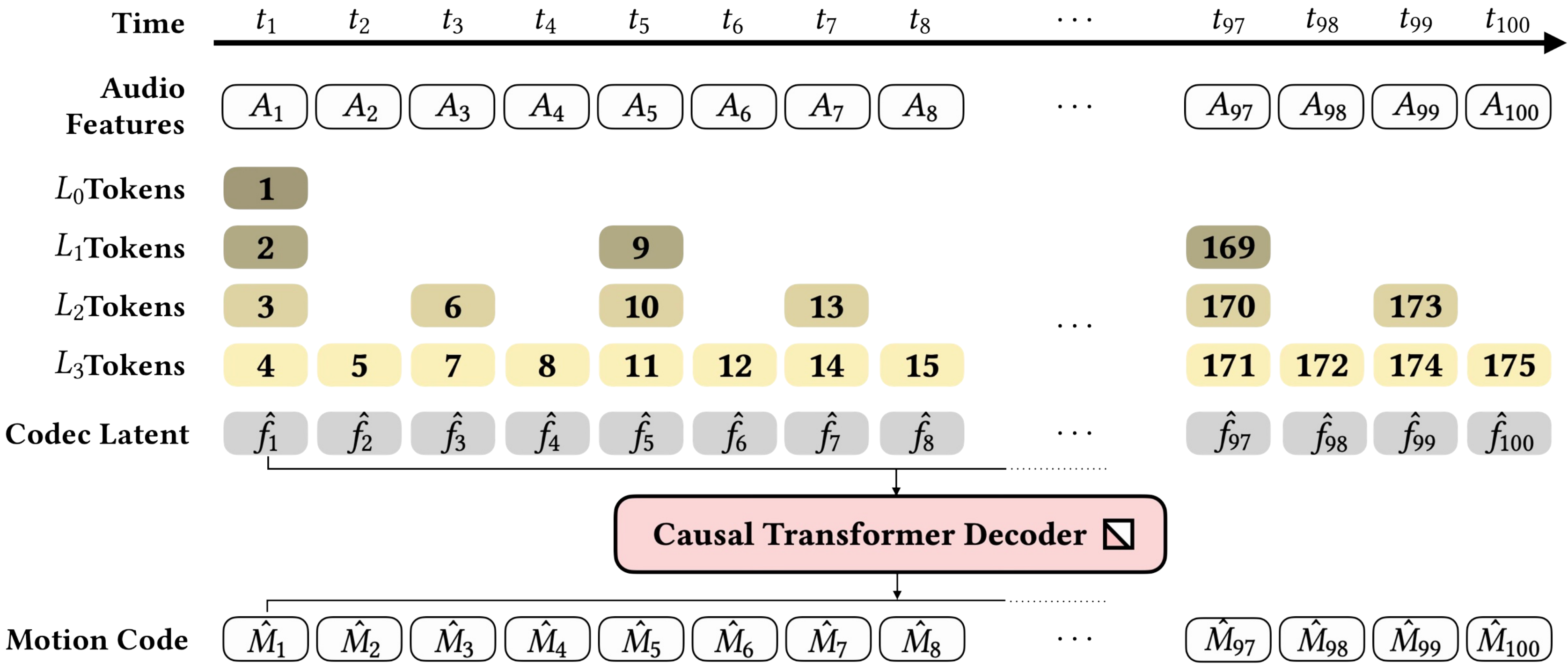}}
\caption{
\textbf{Streaming Multi-resolution Motion Tokens.} Our motion generator predicts hierarchical tokens ($L_0$--$L_3$) frame-by-frame as audio arrives. Each $L_0$ token covers a temporal span of $100$ frames, while tokens at finer levels cover progressively shorter spans. The indexed sequence (0--175) shows the autoregressive order: coarser tokens establish global motion first, followed by interleaved finer tokens for high-frequency detail. This strategy minimizes latency while maintaining multi-scale facial expressivity. The cascading pattern of tokens flowing across resolution levels in time inspires the name \proposed.
}
\label{fig:streaming}
\end{center}
\end{figure}

Following the training of the hierarchical motion codec, we train an autoregressive transformer to map audio into the discrete motion space. We choose autoregressive token prediction over continuous regression or diffusion because it naturally supports frame-by-frame streaming, allows stochastic sampling for motion diversity, and integrates cleanly with the hierarchical token order of \S\ref{sec:32}.
The model operates in a streaming fashion within each temporal window of length $T$, with generation initiated by a learnable start token $S$ at the beginning of each window.
The generator employs cross-attention to fuse three conditional streams: (1) audio embeddings, (2) historical non-tokenized motion from the previous window to ensure temporal continuity, and (3) personalized style priors from our multi-modal retriever to capture identity-specific dynamics.

\paragraph{Streaming Coarse-to-Fine Strategy}
To enable zero-lookahead generation, we propose a Streaming Coarse-to-Fine Strategy that bypasses the conventional requirement of buffering a full audio window.
Instead, generation advances asynchronously: when the latest audio frame $A_t$ still falls within the span of an active coarse code, the model synthesizes only the corresponding fine-grained details; when $A_t$ exceeds that span, the model first predicts a new coarse-level ``anchor'' code to establish the global trend, then fills in the finer scales (Fig.~\ref{fig:streaming}).
This strategy eliminates window-induced latency while preserving the generation diversity inherent in coarse-to-fine autoregressive modeling.

\paragraph{Multi-modal Style Retriever}
We introduce a multi-modal style retriever to resolve the audio-to-motion ambiguity for each identity.
At each time step $t$, the retriever issues a dynamic \textit{multi-modal query} $\mathbf{Q}_t = (\mathbf{A}_t^q, \mathbf{M}_t^q)$, where $\mathbf{A}_t^q$ and $\mathbf{M}_t^q$ are concatenated audio-feature and motion-history windows, respectively. Unlike static or single-modality queries, this joint formulation allows the retriever to disambiguate the search space by simultaneously matching the current audio content and the ongoing motion against the user's characteristics. Specifically, $\mathbf{A}_t^q$ concatenates the previous-window audio features $\mathbf{A}^{\text{gt,prev}} \in \mathbb{R}^{512\times T}$ with the in-window audio seen so far ($A_1, \dots, A_{t-1}$), and $\mathbf{M}_t^q$ is constructed analogously from the previous-window motion $\mathbf{M}^{\text{gt,prev}} \in \mathbb{R}^{137\times T}$.
Crucially, our retriever operates over an unstructured style library rather than a restricted, static set of templates, enabled by the Transformer's content-based addressing across variable-length sequences.

\paragraph{Re-querying Training Strategy}
During training, relying exclusively on ground-truth historical motion $\mathbf{M}_t^q$ for retrieval creates a severe train-inference distribution shift: at inference, retrieval quality degrades once the generated motion deviates from the ground truth, and the output collapses into trivial solutions such as over-smoothed motions that merely replicate the preceding frame $\hat{M}_{t-1}$.
To mitigate this, we propose a Re-querying Training Strategy that introduces a dual-pass optimization.
In the first pass, the model uses $\mathbf{Q}_t = (\mathbf{A}_t^q, \mathbf{M}_t^q)$ as the query to the style library and generates a motion segment $[\hat{M}_1,\dots,\hat{M}_t]$ for $1 \le t \le T$ within the temporal window.
In the second pass, we issue a modified query $\mathbf{Q}'_t = (\mathbf{A}_t^q, \mathbf{M}'^q_t)$, where $\mathbf{M}'^q_t = [\mathbf{M}^{\text{gt,prev}}, \hat{M}_1, \dots, \hat{M}_{t-1}]$ now contains realistic sampling noise from the first pass, and re-query the style library against the synchronized audio.
We denote the re-queried result $\mathbf{P}^\text{prior}_t \in \mathbb{R}^{(T+t)\times 137}$ as a predicted style prior, covering the previous window (length $T$) and the first $t$ positions of the current window.
While $\mathbf{P}^\text{prior}_t$ could in principle live in its own learned latent space, we deliberately constrain it to the original motion space: this (i) avoids the cost of training an auxiliary latent representation, (ii) makes the L2 supervision against $\mathbf{M}^\text{gt}_t = [\mathbf{M}^{\text{gt,prev}}, M_1^\text{gt}, \dots, M_t^\text{gt}]$ directly meaningful in observable motion units, and (iii) ensures that the retrieved priors are directly usable by the generator's cross-attention without an additional decoding step.
By re-querying with predicted (rather than ground-truth) history motion, this dual-pass training forces the retriever to ``recover'' semantically correct stylistic anchors even from imperfect queries. The resulting frame-wise priors give the generator stable guidance and compensate for phonetic ambiguities that would otherwise require future audio lookahead.

\paragraph{Training and Inference}
During training, the autoregressive transformer is supervised in both the discrete code space and the original motion space, to ensure both code accuracy and stylistic alignment. The generation loss $\mathcal{L}_\text{gen}$ is defined as:
\begin{equation}
\label{eq:genloss_all}
\mathcal{L}_\text{gen} = \sum_{l=1}^L \lambda_l \mathcal{L}_\text{ce}(\mathbf{\hat{C}}_l, \mathbf{C}_l^\text{gt}) + \frac{\lambda_p}{T}\sum_{t=1}^{T} \left\| \mathbf{P}^\text{prior}_t - \mathbf{M}^\text{gt}_t \right\|^2,
\end{equation}
where $\mathcal{L}_\text{ce}$ is the cross-entropy loss between the predicted code $\hat{C}_l$ and the ground-truth codes $C_l^{\text{gt}} = [E(\mathbf{M}^\text{gt})]_l$ at each hierarchical level $l$ with level-specific weights $\lambda_l$, and $E$ is the temporal hierarchical encoder pre-trained in \S\ref{sec:32}.
At inference, we sample from the predicted logit distributions (rather than greedy-decode) for diverse motion synthesis. We further apply Classifier-Free Guidance (CFG)~\citep{ho2022cfg}: each of the three conditioning streams (history motion, audio, retrieved style) is independently dropped at training, and a guidance scale re-weights conditioned and unconditioned logits at inference, balancing motion expressivity against alignment to audio and the style prior.

\section{Experiments}
\subsection{Experimental Settings}
\paragraph{Datasets}
To assess performance and generalization to public benchmarks, we evaluate our framework on two datasets.
\textit{Internal dataset.} For our main comparison (Table~\ref{tab:main_exp}), we use an internal dataset of $34{,}906$ video clips from $832$ subjects ($469.6$ hours) captured in a controlled environment, encompassing speech, non-verbal vocalizations, and silent gestures to provide diverse but consistent identity signatures.
We partition the data into $784$ identities for training and $48$ for testing, and extract per-frame motions using the pre-trained imitator models from~\citep{agrawal2025seamlessinteractiondyadicaudiovisual}, which can be visualized by a paired 2D renderer capable of identity-conditioned synthesis via a single reference image.
\textit{Public dataset (FLAME-based).} To ensure a fair comparison and verify that our gains generalize beyond the internal data distribution, we additionally re-train and evaluate on the TalkVid-Core dataset~\citep{talkvid2025} (Table~\ref{tab:main_exp_flame}), applying additional filtering to retain clean, speaker-centric clips.
The resulting split contains $12{,}626$ training clips (140h 28m, $2{,}115$ unique identities) and $485$ test clips (5h 06m, $80$ identities); here motions are represented using the FLAME parametric head model~\citep{li2017learning}.

\paragraph{Implementation Details}
All experiments were conducted on 8 NVIDIA A100 GPUs using a two-stage training strategy.
In the first stage, we train the multi-scale codec for $37,500$ steps with a total batch size of $512$.
In the second stage, the codec weights are frozen, and the transformer-based generator is trained for $37,500$ steps with the same batch size.
Both stages utilize the Adam optimizer with a base learning rate of $8 \times 10^{-4}$.
Training the first stage requires approximately 20 GPU hours (2.5 hours wall-clock with 8 GPUs), while the second stage requires roughly 90 GPU hours (11 hours wall-clock with 8 GPUs).
Please refer to the supplementary materials for more details on model structure and hyperparameters.

\paragraph{Runtime.}
On an NVIDIA H100 GPU, our model takes $24.8$\,ms per frame ($40.3$ FPS) and has an end-to-end latency of $104.8$\,ms, including the $80$\,ms audio input required by the encoder. Increasing the reference library from $1\times200$ to $5\times200$ frames raises retrieval time by only $13\%$ ($574$ to $649$\,ms per $100$ generated frames). Full runtime and memory statistics are provided in the supplementary material.

\subsection{Quantitative Results}

\begin{figure}
\begin{center}
\centerline{\includegraphics[width=1.0\columnwidth]{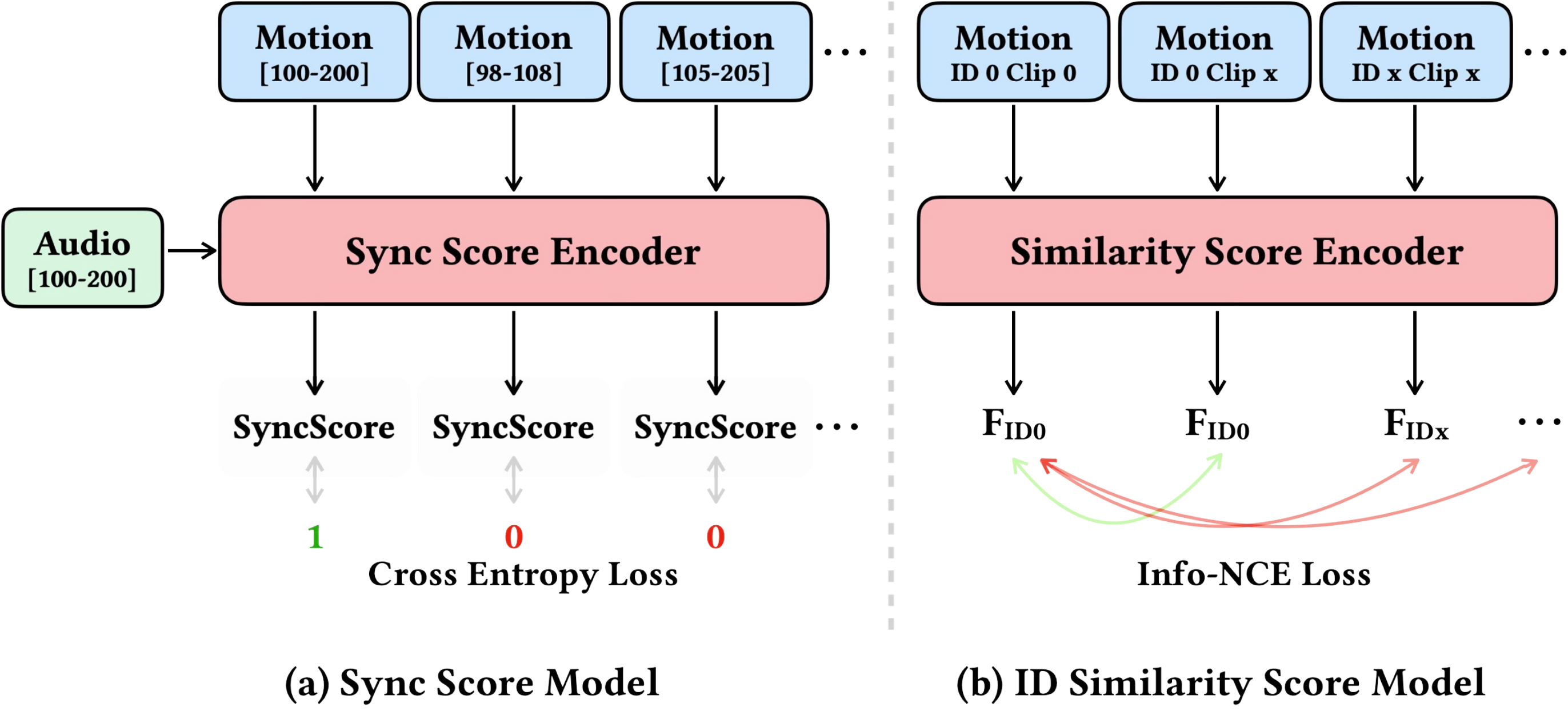}}
\caption{
\textbf{Training Objectives for Evaluation Metrics.}
(a) Synchronization: The model is trained via binary cross-entropy; pairs with an audio-motion offset under 2 frames are positive, others are negative.
(b) Identity Consistency: InfoNCE loss pulls same-identity embeddings together (green line) and pushes differing identities apart (red line) in the feature space.
}

\label{fig:eval_metrics}
\end{center}
\Description{Training objectives for the Sync and Similarity evaluators: a binary cross-entropy classifier for audio-motion synchronization on the left, and an InfoNCE contrastive loss for identity similarity on the right.}
\end{figure}

\begin{table*}[t]
\caption{
    \textbf{Quantitative results.} ``Streaming'' denotes zero-lookahead generation. Partial personalization ({\hcheck}) uses static IDs (ARTalk and DiffPoseTalk) or audio-only history (MemoryTalker), whereas ours jointly uses audio and motion. Best and second-best results are \au{gold} and \ag{silver}.
}
\label{tab:main_exp}
\tablestyle{10.5pt}{1.05}
\begin{center}
\begin{tabular*}{\linewidth}{@{\extracolsep{\fill}}l|cc|cccc}
\toprule[1.25pt]
Method & Streaming & Personalized & Sync Score $\uparrow$ & Sim Score $\uparrow$ & ID-FED $\downarrow$ & ID-FPD $\downarrow$ \\
\midrule
DiffPoseTalk \citep{sun2024diffposetalk}     & \rcross   & \hcheck   & 0.468 & 0.525 & \ag{27.38} & \ag{3.57} \\
ARTalk \citep{chu2024artalk}                 & \rcross   & \hcheck   & \ag{0.479} & \ag{0.596} & 31.65 & 3.88 \\
AudioRTA \citep{lee2025audio}                & \gcheck   & \rcross   & 0.329 & 0.165 & 46.33 & 3.75 \\
MemoryTalker \citep{kim2025memorytalker}     & \rcross   & \hcheck   & 0.345 & 0.117 & 67.85 & 6.58 \\
\midrule
Ours                                        & \gcheck   & \gcheck   & \au{0.509} & \au{0.718} & \au{17.58} & \au{2.07} \\
\bottomrule[1.25pt]
\end{tabular*}
\end{center}
\end{table*}
\paragraph{Evaluation Metrics}
On our internal imitator-based dataset (Table~\ref{tab:main_exp}), motion is rendered by a 2D decoder rather than a 3D mesh, so we report four latent-space metrics. \textbf{ID-FED} and \textbf{ID-FPD} are ID-wise Fr\'echet distances on expression and pose, computed per identity and averaged across identities to better verify identity-specific behavior. \textbf{Sync Score} and \textbf{Similarity Score} are produced by two specialized evaluation models (Fig.~\ref{fig:eval_metrics}): the former measures audio-motion alignment via binary classification against temporally perturbed negatives; the latter measures ID stylistic consistency as cosine similarity in an InfoNCE-trained identity embedding space. Architecture and training details of both evaluators are in the supplementary material.
On the public FLAME-based benchmark (Table~\ref{tab:main_exp_flame}), the 3D mesh is directly available, so we additionally adopt standard FLAME-domain metrics following prior work~\citep{fan2022faceformer, nocentini2024scantalk, richard2021meshtalk, sun2024diffposetalk, xing2023codetalker, chu2024artalk}: \textbf{LVE}~\citep{richard2021meshtalk} is the per-frame maximum $L_2$ error over lip vertices and \textbf{FDD}~\citep{xing2023codetalker} is the temporal standard deviation of upper-face vertex motion against ground truth. Sync Score, Similarity Score, ID-FED, and ID-FPD are computed analogously to the internal-dataset setting.

Table~\ref{tab:main_exp} presents the quantitative comparison of our method against state-of-the-art approaches, averaged across all testing clips.
Our method achieves superior performance across all metrics.
Specifically, the improvements in Sync Score and Similarity Score demonstrate our model's ability to not only synchronize lip movements accurately but also capture the unique speaking style of the target identity.
To address concerns about evaluation on our internal dataset, we additionally re-train and evaluate our framework on a public FLAME-based benchmark. As shown in Table~\ref{tab:main_exp_flame}, our method maintains its lead on the standard FLAME-domain metrics (LVE, FDD) as well as on the identity-specific metrics (Sim Score, ID-FED, ID-FPD), confirming that the gains generalize beyond the imitator representation and our internal data distribution.

\begin{table}[t]
\caption{
    \textbf{Quantitative results on a public FLAME-based benchmark.}
    First-best results are highlighted in \au{gold}.
    ID-FPD for FLAME is scaled by 1000×.
    ID-FPD is not reported for CodeTalker or TalkingStyle because these methods do not generate head motion.
}
\label{tab:main_exp_flame}
\tablestyle{2.5pt}{1.05}
\begin{center}
\begin{tabular*}{\linewidth}{@{\extracolsep{\fill}}lcccccc}
\toprule[1.25pt]
Method & LVE $\downarrow$ & FDD $\downarrow$ & Sync $\uparrow$ & Sim $\uparrow$ & ID-FED $\downarrow$ & ID-FPD $\downarrow$ \\
\midrule
CodeTalker   & 12.23 & 24.21 & 0.401 & 0.332 & 21.35 & -    \\
TalkingStyle & 11.72 & 23.08 & 0.413 & 0.310 & 20.72 & -    \\
DiffPoseTalk               & 9.35  & 20.80 & 0.496 & 0.564 & 15.28 & 2.29 \\
ARTalk                     & 8.81  & 18.50 & 0.523 & 0.601 & 13.01 & 0.95 \\
\midrule
Ours          & \au{8.03} & \au{13.24} & \au{0.574} & \au{0.678} & \au{9.30} & \au{0.60} \\
\bottomrule[1.25pt]
\end{tabular*}
\end{center}
\end{table}

\subsection{Qualitative Results}
\begin{figure*}[t]
\begin{center}
\centerline{\includegraphics[width=0.85\linewidth]{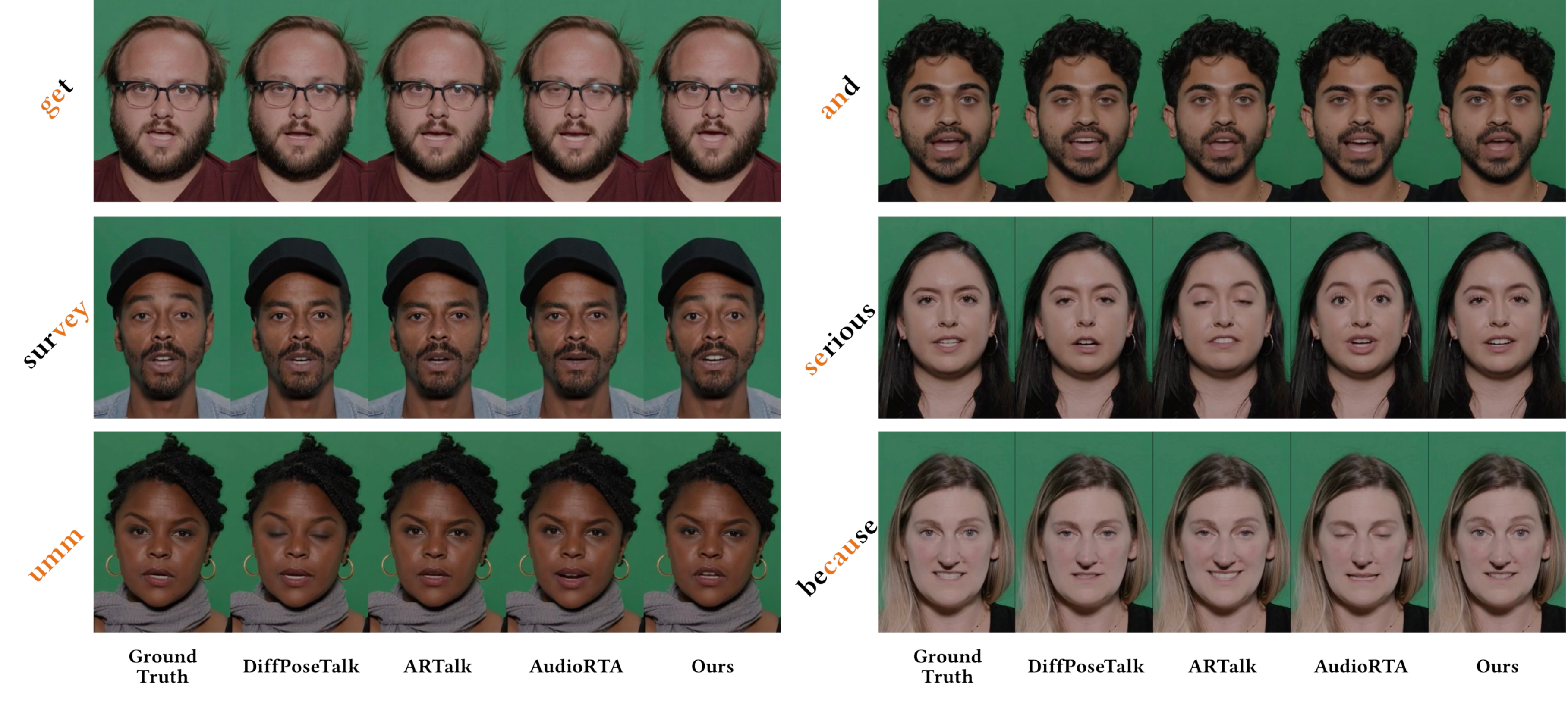}}
\caption{
\textbf{Qualitative Comparison of Phoneme Articulation.} We compare lip shape synthesis across distinct phonemes against various baselines. Our method produces lip configurations that more closely align with the ground truth. See the supplementary video for the full comparison. All identities shown are synthetic.
}
\label{fig:qualitative_1}
\end{center}
\Description{A grid of synthesized face frames showing each baseline and our method articulating the same set of distinct phonemes, with the ground-truth mouth shape in the rightmost column. Our method's lip configurations are visibly closer to the ground truth than the baselines'.}
\end{figure*}

\begin{figure*}[t]
\begin{center}
\centerline{\includegraphics[width=0.85\linewidth]{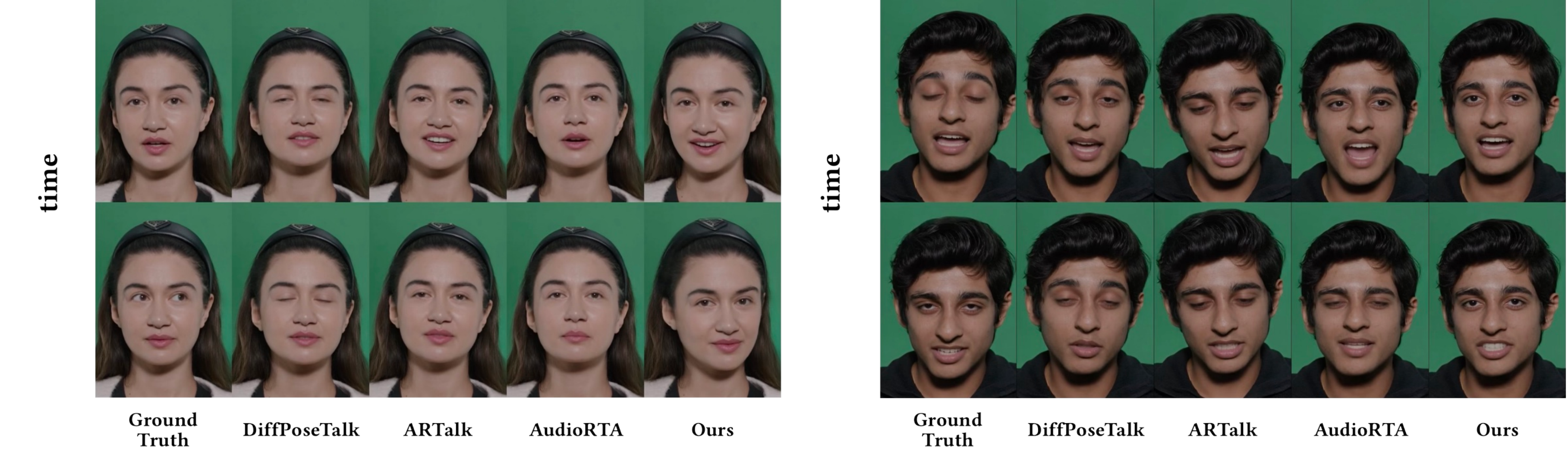}}
\caption{
\textbf{Qualitative Head Pose Comparison.} We compare head poses at two temporally distant frames within a single sequence. Baseline methods exhibit restricted movement, whereas our model generates significantly more diverse and realistic motion trajectories. All identities shown are synthetic.
}
\label{fig:qualitative_2}
\end{center}
\Description{A side-by-side grid showing two temporally distant frames from each baseline and our method on the same audio clip. Baseline frames look nearly identical in head pose, while our method's two frames show distinctly different, more natural head orientations.}
\end{figure*}

Our model produces better lip-sync and more diverse head motion compared to the baselines, as evidenced by Fig.~\ref{fig:qualitative_1} and Fig.~\ref{fig:qualitative_2}. In Fig.~\ref{fig:qualitative_1}, we compare our results with those of the baselines on pronouncing different phonemes. In Fig.~\ref{fig:qualitative_2}, we show two frames of each video to demonstrate the head movements across time. Full videos are provided in the supplementary material.

\setcounter{dbltopnumber}{2}

\subsection{User Study}

\begin{table}[t]
\caption{
    \textbf{User Study Results.}
    Percentage of participants who \textit{preferred our method over each baseline} on three criterias with $95\%$ Wilson confidence intervals in brackets ($n \approx 112$ A/B votes per cell). Lip Sync (temporal audio-lip alignment), ID Sim (consistency of facial style and identity), and Naturalness (fluidity of expressions and head poses). 
}
\label{tab:user_study}
\tablestyle{2.5pt}{1.1}
\begin{center}
\begin{tabular*}{\linewidth}{@{\extracolsep{\fill}}lccc}
\toprule[1.25pt]
Comparison & Lip Sync $\uparrow$ & ID Sim $\uparrow$ & Naturalness $\uparrow$ \\
\midrule
ARTalk        & 68.5\%~{\scriptsize[59.4, 76.4]} & 76.8\%~{\scriptsize[68.2, 83.6]} & 86.5\%~{\scriptsize[78.9, 91.6]} \\
DiffPoseTalk  & 67.9\%~{\scriptsize[58.8, 75.8]} & 68.8\%~{\scriptsize[59.7, 76.6]} & 61.6\%~{\scriptsize[52.4, 70.1]} \\
AudioRTA      & 78.6\%~{\scriptsize[70.1, 85.2]} & 71.4\%~{\scriptsize[62.4, 79.0]} & 86.6\%~{\scriptsize[79.1, 91.7]} \\
MemoryTalker  & 86.9\%~{\scriptsize[79.4, 91.9]} & 73.8\%~{\scriptsize[65.0, 81.1]} & 85.5\%~{\scriptsize[77.8, 90.8]} \\
\bottomrule[1.25pt]
\end{tabular*}
\end{center}
\end{table}

To complement our objective metrics and assess the perceptual quality of the generated videos, we conducted a user study with $15$ general (non-expert) participants. Each participant viewed $30$ side-by-side video pairs in randomized order, with each pair contrasting our method against one of the four baselines on three criteria: lip synchronization (alignment of mouth movements with audio), ID similarity (fidelity of unique behavioral traits to the target ID), and visual naturalness (realism of facial expressions and head motion), yielding $450$ A/B comparisons in total.
As shown in Table~\ref{tab:user_study}, participants preferred our method over every baseline across all three criteria, with the strongest margins over MemoryTalker (e.g., $86.9\%$ on Lip Sync, $85.5\%$ on Naturalness) and the narrowest---but still majority---margin over DiffPoseTalk on Naturalness ($61.6\%$), consistent with DiffPoseTalk being the closest competitor on visual quality in Table~\ref{tab:main_exp}.

\subsection{Ablation Study}

\begin{table}[t]
\caption{
    \textbf{Ablation Study Results.} We evaluate the impact of the hierarchical motion codec (HMC), multi-modal retrieval, and retrieval frequency (per-frame querying versus a ``lazy-update'' baseline that re-queries only once every $T$ frames), and validate the re-query mechanism, which prevents the Sync-Score collapse caused by train--inference exposure bias. First and second-best results are highlighted in \au{gold} and \ag{silver}.
}
\label{tab:ablation_exp}
\tablestyle{4.5pt}{1.05}
\begin{center}
\begin{tabular*}{\linewidth}{@{\extracolsep{\fill}}lcccc}
\toprule[1.25pt]
Method & Sync $\uparrow$ & Sim $\uparrow$ & ID-FED $\downarrow$ & ID-FPD $\downarrow$ \\
\midrule
w/o HMC               & 0.304 & 0.505 & 37.88 & 3.59 \\
w/o Retrieval         & 0.439 & 0.120 & 52.60 & 3.94 \\
Audio-only Retrieval  & 0.407 & \au{0.779} & \ag{18.96} & 2.36 \\
Motion-only Retrieval & \ag{0.509} & 0.627 & 41.92 & 9.07 \\
Lazy-update Retrieval & 0.436 & 0.708 & 20.35 & 2.14 \\
w/o Re-query          & 0.304 & 0.617 & 26.01 & 3.10 \\
\midrule
Full (Lookahead 0)    & \ag{0.509} & 0.718 & \au{17.58} & \ag{2.07} \\
Full (Lookahead 5)    & \au{0.625} & \ag{0.725} & 19.57 & \au{1.95} \\

\bottomrule[1.25pt]
\end{tabular*}
\end{center}
\end{table}

\begin{figure}
\begin{center}
\centerline{\includegraphics[width=1.0\columnwidth]{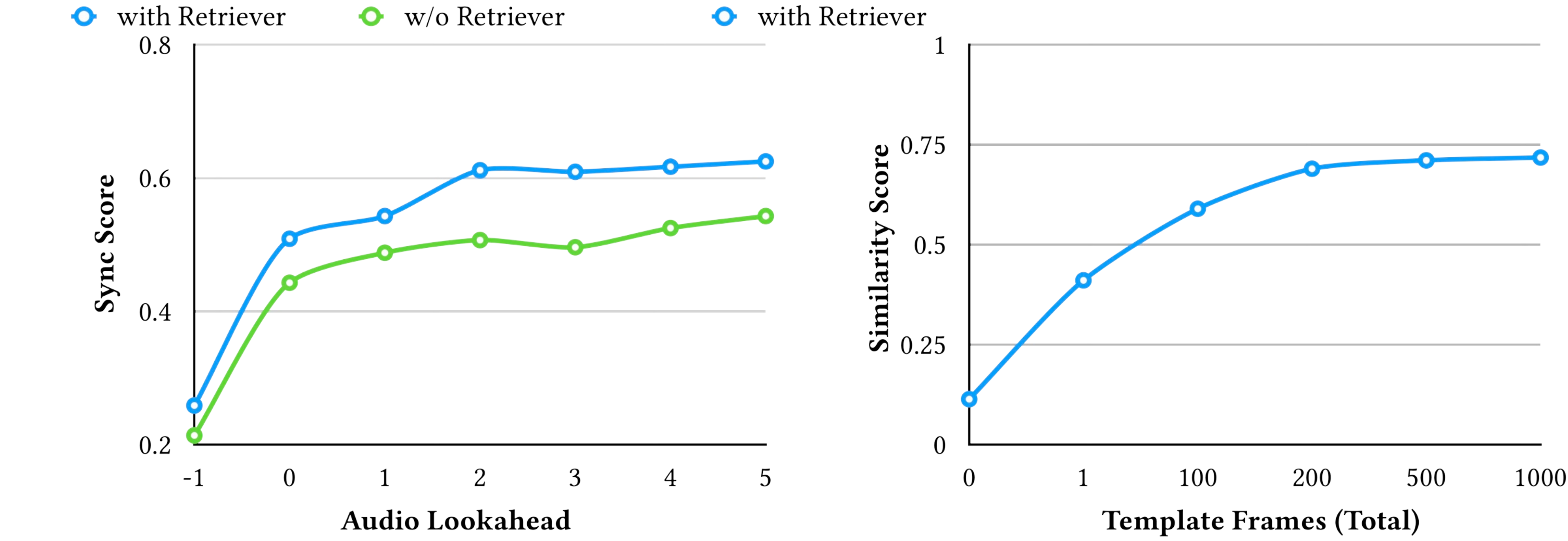}}
\caption{
    \textbf{Ablation Studies.} (a) \textbf{Retrieval Impact on Synchronization}: our retrieval mechanism consistently improves lip-sync quality across the lookahead settings $k \in \{0, 1, 2, 3, 5\}$. (b) \textbf{Library Scalability}: performance gains scale monotonically with the size of the reference library, without requiring model retraining.
}
\label{fig:ablation_exp}
\end{center}
\end{figure}

\paragraph{Hierarchical Motion Codec.}
We ablate the hierarchical multi-scale modeling by replacing our codec with a single-layer causal variant. As shown in Table~\ref{tab:ablation_exp} (``w/o HMC''), generation quality significantly declines: Sync drops from $0.509$ to $0.304$ and ID-FED nearly doubles. We attribute this to the hierarchical architecture's ability to progressively refine details through inter-layer autoregression; the multi-layer token space also supports stochastic sampling that introduces richer motion diversity.

\paragraph{Effect of Multi-modal Style Retrieval.}
We compare our full retriever against three variants: no retriever, audio-only retriever, and motion-only retriever. Removing the retriever has minimal impact on lip-sync (since Sync depends primarily on the audio input) but causes a large drop in ID similarity (Sim $0.120$ vs $0.718$, ID-FED $52.60$ vs $17.58$): without identity priors, the generator falls back to an ``average'' speaker. The single-modality variants reveal complementary failure modes. The motion-only retriever matches our full Sync ($0.509$) because audio still drives the generator's lip-sync head directly, but its ID-FPD nearly quadruples ($9.07$ vs $2.07$) since it cannot disambiguate library clips by audio context. The audio-only retriever, conversely, struggles in non-verbal regions where audio is uninformative, yielding a worse ID-FPD ($2.36$ vs $2.07$) despite an apparently strong Sim Score. Our full retriever achieves the most balanced performance by jointly conditioning on audio and motion history.

\paragraph{Effect of Re-query Training Strategy.}
We compare our full model against a version trained without re-query. Table~\ref{tab:ablation_exp} shows that removing re-query causes severe collapse: Sync drops from $0.509$ to $0.304$, on par with removing the hierarchical codec entirely. This may appear surprising since re-query targets retrieval robustness rather than audio-lip alignment, but the failure is a chain of effects. At training time, the retriever sees only ground-truth motion history; at inference, it receives the generator's own predictions, which contain sampling noise. This distribution shift causes retrieval to collapse, the generator loses ID-specific guidance, and the output reverts to an ``average mouth shape'' that the Sync evaluator penalizes. Re-query training closes this loop by exposing the retriever to predicted (noisy) motion at training time, forcing it to recover the correct stylistic features even from imperfect queries.

\paragraph{Effect of Audio Lookahead in Streaming.}
While zero-lookahead minimizes latency, we also investigate the gains from a small future-audio window. ``Lookahead $k$'' denotes that at time $t$ the model can access audio features up to $t+k$. As Table~\ref{tab:ablation_exp} shows, increasing the lookahead (e.g., $k=5$) notably improves lip-sync ($0.509 \to 0.625$): the model leverages upcoming speech to synthesize more stable and physiologically plausible lip motion. Our framework nevertheless remains competitive under the strict $k=0$ constraint, supporting real-time, zero-lookahead streaming. Qualitative results are in the supplementary video.

\paragraph{Effect of Number of Templates.}
Templates are randomly sampled at training and inference. In Fig.~\ref{fig:ablation_exp}(b), $1\times100$ denotes one $100$-frame sequence and $5\times200$ five $200$-frame sequences. Increasing the template count consistently improves ID similarity by exposing the retriever to broader conversational and emotional contexts. However, gains plateau beyond a small threshold, indicating that the model distills the core stylistic traits from a few representative samples; high-fidelity stylization is therefore achievable with only a handful of short clips. Qualitative results are in the supplementary video.

\section{Conclusion}
We present \proposed, an end-to-end causal framework for low-latency audio-driven facial motion synthesis. A streaming coarse-to-fine strategy eliminates window-induced buffering delays for true frame-by-frame generation, while a multi-modal style retriever resolves audio-to-motion ambiguity via identity-specific reference clips. Experiments on both an internal imitator-based dataset and a public FLAME-based benchmark show that \proposed outperforms state-of-the-art baselines in lip-sync accuracy and identity preservation while supporting zero-lookahead streaming at real-time throughput.

While robust, our framework exhibits slight latency at silence-to-speech transitions, and stylization quality depends on the richness of input templates: sparse libraries may yield generic motion. Future work will explore generative priors to reduce template dependence and extend the streaming architecture to full-body gesture synthesis for more immersive holographic telepresence.

\bibliographystyle{ACM-Reference-Format}
\bibliography{references}

\appendix
\clearpage
\newcommand{\ua}{$\uparrow$}
\newcommand{\da}{$\downarrow$}
\definecolor{lightgreen}{HTML}{D8ECD1}
\newcommand{\bt}[1]{\colorbox{lightgreen}{#1}}

\section{Implementation Details}
\label{sec:supp_impl}

We provide complete architecture and training specifications for the three trained models (the hierarchical motion codec, the streaming motion generator, and the style retriever), the two evaluation models that produce our Sync and Similarity scores, and our data pipeline. All models are implemented in PyTorch and trained on 8 NVIDIA A100 GPUs using Adam (default $\beta_1 = 0.9$, $\beta_2 = 0.999$).

\subsection{Hierarchical Motion Codec}
\label{sec:supp_codec}

\paragraph{Architecture.}
The motion sequence is processed in non-overlapping windows of $T = 100$ frames (4\,s at $25$\,Hz). 
The encoder and the decoder are both pre-LN Transformers with $L_\text{tx} = 6$ blocks, hidden dimension $d = 384$, $6$ attention heads, an MLP expansion of $1.5\times$ (hidden width $576$), GELU(tanh) activation, no dropout, and RoPE positional encoding (maximum sequence length $240$). 
The decoder applies a triangular causal attention mask at every layer, whereas the encoder uses bidirectional attention. The decoder upsamples codes using a floor-indexed nearest-neighbor operator that reads only from past positions.

\paragraph{Hierarchical quantization.}
BSQ~\citep{zhao2024imagevideotokenizationbinary} produces $p = 32$-bit binary codes (implicit codebook size $2^{32}$) at each of $L = 4$ hierarchical levels with code lengths $r_1 = 1$, $r_2 = 25$, $r_3 = 50$, $r_4 = 100$ (coarse to fine). Residual quantization accumulates the dequantized codes across levels; BSQ's inverse-temperature factor is set to $100$. The entropy regularizer follows BSQ's two-term formulation, namely the per-sample binary entropy minus the codebook-marginal binary entropy.

\paragraph{Loss weights.}
The $L_1$ reconstruction term uses weight $1.0$.
We additionally add a per-channel $L_1$ term on the pose subset $R_i$ (weight $8.0$) so that pose accuracy is not dominated by the much larger expression subspace with FLAME version.
The pose-only smoothness terms use $\lambda_v = 5.0$ (velocity) and $\lambda_j = 0.5$ (jitter). 
The entropy regularizer term of BSQ is set to $\lambda_e = 1.0$.

\paragraph{Training.}
We use Adam with a base learning rate of $8 \times 10^{-4}$, a $1{,}000$-step linear warm-up, and linear decay to $8 \times 10^{-5}$ over $37{,}500$ total optimizer steps; the total batch size is $512$ (distributed as $64$ per GPU across 8 GPUs), and the EMA decay is $0.9999$, updated every step.
Training takes approximately $20$ GPU-hours (about $2.5$\,h wall-clock on 8 GPUs).

\subsection{Streaming Motion Generator and Style Retriever}
\label{sec:supp_qgen}

\paragraph{Generator architecture.}
The autoregressive Transformer has $6$ pre-LN blocks with hidden dimension $384$, $6$ attention heads, and MLP expansion $4\times$ (hidden width $1{,}536$). We apply stochastic depth that increases linearly from $0$ to $0.025$ across blocks. Each block fuses three conditioning streams via cross-attention: audio embeddings $A_i \in \mathbb{R}^{512}$ from the pre-trained Mimi backbone~\citep{defossez2024moshi}, the previous-window motion code (raw dimension $p = 32$ per BSQ token), and the retrieved style prior $\mathbf{P}^\text{prior}_t \in \mathbb{R}^{(T+t)\times 137}$ from the style retriever. All three streams are linearly projected to the generator's embedding dimension $384$ before cross-attention.

\paragraph{Style Retriever.}
The retriever is a $4$-block Transformer with embedding dimension $256$, $4$ attention heads, MLP expansion $4\times$, and RoPE positional encoding. It ingests a joint query of dimension $d_\text{audio} + d_\text{motion} = 512 + 137 = 649$ over the user's reference clips; keys are linearly projected to $256$, and values are the raw motion vectors (dimension $137$).

\paragraph{Re-querying schedule.}
The dual-pass re-query is applied at every training step. The second-pass query is constructed from the first pass's predicted motion tokens; for sampling we mix top-$p$ ($p = 0.97$) with probability $0.5$ and greedy (argmax) with probability $0.5$, so the retriever sees a range of noise levels in its query.

\paragraph{Classifier-free guidance.}
The history-motion, audio, and retrieved-style streams are each independently dropped with probability $p_\text{cfg} = 0.1$ during training. At inference we sample with guidance scale $w = 2.0$, temperature $\tau = 1.0$, and nucleus $p = 0.97$.

\paragraph{Reference library construction.}
During training, we randomly crop one $200$-frame motion segment from an arbitrary clip of the target identity, using a clip different from the current training sample. At inference, the reference library may contain one or more motion segments, randomly cropped from arbitrary clips of the target identity; the segment lengths and counts follow the setting of each experiment (e.g., $1\times200$ or $5\times200$). We apply no deduplication or quality-based filtering to the reference segments.

\paragraph{Loss weights.}
Per-level cross-entropy uses $\lambda_l = 1.0$ for every hierarchy level. The re-query MSE term against the ground-truth motion uses $\lambda_p = 1.0$.

\paragraph{Training.}
With the codec weights frozen, we train the generator using Adam with the same schedule as the codec: base learning rate $8 \times 10^{-4}$, $1{,}000$-step linear warm-up, and linear decay to $8 \times 10^{-5}$ over $37{,}500$ total optimizer steps; total batch size $512$ ($64$ per GPU across 8 GPUs); EMA decay $0.9999$. Training takes approximately $90$ GPU-hours (about $11$\,h wall-clock on 8 GPUs).

\subsection{Evaluation Models}
\label{sec:supp_evals}

The Sync Score and Similarity Score are produced by two small discriminator networks.

\paragraph{Sync Evaluator.}
A $4$-block Transformer with hidden dimension $256$, $4$ attention heads, and a $128$-D feature dimension. It ingests a $T = 100$-frame motion-code window (projected from the $32$-D BSQ tokens to $256$) and cross-attends to the matched $4$-second Mimi audio embedding stream (projected from $512$ to $256$). A LayerNorm and a linear head map the output to a per-frame binary logit. The training objective is binary cross-entropy over three pair types. Positives are motion--audio pairs aligned within $\pm 1$ frame. Hard negatives are time-shifted by $\pm 2$ to $\pm 3$ frames and contribute through a separate cross-entropy term of equal weight to the easy term. Easy negatives are time-shifted by $\pm 5$ to $\pm 50$ frames or are sampled from a different identity.
Training uses Adam with learning rate $8 \times 10^{-4}$, total batch size $512$ ($64$ per GPU across 8 GPUs), $12{,}500$ total optimizer steps, and EMA decay $0.9999$.

\paragraph{Similarity Evaluator.}
A motion-only $4$-block Transformer of the same width as the Sync Evaluator (hidden $256$, $4$ heads, $128$-D feature). It encodes a $T = 100$-frame motion window, mean-pools across time, and projects to a $137$-D identity embedding which is then $L_2$-normalized. Training uses InfoNCE at temperature $\tau = 0.5$: positives are pairs from the same Person ID, negatives are from different IDs, and intra-clip neighbors of the same Person ID are down-weighted to $0.1$ to discourage trivial positives. Training uses Adam with learning rate $1 \times 10^{-4}$, total batch size $512$ ($64$ per GPU across 8 GPUs), $4{,}000$ total optimizer steps, and EMA decay $0.9999$.

\paragraph{Computing the reported metrics.}
At evaluation time, the Sync Evaluator's softmax positive-class probability is averaged across frames within a clip to produce the per-clip Sync Score, then averaged across the test set. The Similarity Evaluator's cosine similarity between predicted and ground-truth identity embeddings is averaged across clips within each identity and then across identities, producing the ID-wise Similarity Score.

\paragraph{Validation on held-out identities.}
The evaluators use the same identity-disjoint training and test split as the main model. On the held-out TalkVid test set, the Similarity Evaluator produces an average score of $0.8404$ for same-identity pairs and $0.2998$ for different-identity pairs. The Sync Evaluator is likewise sensitive to temporal misalignment: its average score decreases from $0.8276$ for exactly aligned audio--motion pairs to $0.7280$ and $0.2624$ after shifts of one and two frames, respectively. These results show that the learned evaluators distinguish identity similarity and audio--motion synchronization on unseen identities. Their conclusions are also consistent with the standard LVE and FDD metrics reported on TalkVid.

\subsection{Data Processing}
\label{sec:supp_data}

\paragraph{Baseline protocol.}
All baselines are retrained using their official open-source implementations on the same internal dataset and TalkVid split used by our method. To align each baseline with the motion representation used for evaluation, we modify only its final output layer to predict the corresponding per-frame motion vector: $137$ dimensions on the internal dataset and $108$ dimensions on TalkVid. All methods are then evaluated using the same inference protocol and metrics.

\paragraph{Audio.}
Raw audio is resampled to $24$\,kHz, right-padded so its length is a multiple of one Mimi frame, and encoded by the pre-trained Mimi backbone. The resulting features are upsampled by $2\times$ via nearest-neighbor interpolation to align with the $25$\,Hz motion frame rate, yielding $A_i \in \mathbb{R}^{512}$ time-aligned with the motion $M_i$.

\paragraph{Filtering.}
For TalkVid-Core, we discard videos whose average head orientation deviates by more than $40^\circ$ from the frontal direction, because the FLAME extractor is unreliable on profile views and can introduce contaminated motion estimates. We then construct an identity-disjoint split by randomly holding out $80$ identities for testing and using all remaining identities for training. No additional silence removal or face-detection thresholding is applied at the dataloader level. The internal imitator-based dataset is captured in a controlled environment and therefore requires no further filtering. For the public-benchmark experiments, motion is represented by a $108$-D FLAME motion vector~\citep{li2017learning}, consisting of $100$ expression and $8$ pose dimensions; all other hyperparameters above remain unchanged.

\paragraph{FLAME tracking on TalkVid-Core.}
For the public-benchmark experiments, both our method and DiffPoseTalk~\citep{sun2024diffposetalk} operate in the FLAME parameter space, so we extract FLAME~\citep{li2017learning} parameters from the TalkVid-Core~\citep{talkvid2025} videos as a preprocessing step. We resample each clip to $25$\,fps and use EMICA~\citep{DECA2021, MICA2022, EMOCA2021, gpavatar2024, chu2024gagavatar} to extract a $100$-dimensional shape vector, a $100$-dimensional expression vector, and a $8$-dimensional pose vector per frame.

\paragraph{Rendering protocol.}
For 2D visualization, we use a third-party pretrained renderer that takes an identity image and a $137$-D motion representation as input and produces expression images. The renderer is used only for visualization, and no quantitative evaluation depends on its output. All compared methods use the same renderer and weights, so the rendering process does not affect comparison fairness. For 3D visualization, we use PyTorch3D following the rendering setup of ARTalk~\citep{chu2024artalk}.

\section{Runtime Analysis}
\label{sec:supp_runtime}

We profile the streaming pipeline on an NVIDIA H100 GPU using TalkVid, with a reference library of $1\times200$ frames and a target motion rate of $25$ FPS. Table~\ref{tab:supp_runtime_overall} summarizes the overall performance. The measured end-to-end latency comprises $24.8$\,ms of model inference and $80$\,ms to collect the minimum audio input required by the audio encoder. This input-collection interval does not introduce future-audio conditioning into the motion generator, which remains strictly zero-lookahead.

\begin{table}[t]
\caption{Overall streaming performance on an NVIDIA H100 GPU.}
\label{tab:supp_runtime_overall}
\tablestyle{8pt}{1.10}
\begin{center}
\begin{tabular*}{\columnwidth}{@{\extracolsep{\fill}}lr}
\toprule[1.25pt]
Metric & Result \\
\midrule
Per-frame inference time & $24.8$ ms \\
Sustained throughput & $40.3$ FPS \\
Audio input collection & $80.0$ ms \\
End-to-end latency & $104.8$ ms \\
Real-time factor & $0.620$ \\
Peak GPU memory & $3.86$ GB \\
\bottomrule[1.25pt]
\end{tabular*}
\end{center}
\end{table}

Over $100$ generated frames, corresponding to $4$ seconds of motion, the model requires $2.481$ seconds of computation. Table~\ref{tab:supp_runtime_breakdown} reports the component-wise breakdown. The autoregressive transformer is the largest contributor ($44.5\%$), while retrieval accounts for $23.1\%$ of the total runtime. Increasing the reference library from $1\times200$ to $5\times200$ frames increases retriever time by only $13\%$ ($574$\,ms to $649$\,ms), with a small impact on the overall inference cost because the autoregressive sequence length remains constant and the additional computation arises primarily from cross-attention over the larger library.

\begin{table}[t]
\caption{Component-wise runtime for $100$ generated frames, measured on an NVIDIA H100 GPU.}
\label{tab:supp_runtime_breakdown}
\tablestyle{8pt}{1.10}
\begin{center}
\begin{tabular*}{\columnwidth}{@{\extracolsep{\fill}}lrr}
\toprule[1.25pt]
Component & Time (ms) & Fraction \\
\midrule
Audio encoder & $202$ & $8.1\%$ \\
Transformer & $1{,}105$ & $44.5\%$ \\
Retriever & $574$ & $23.1\%$ \\
Decoding & $600$ & $24.2\%$ \\
\midrule
Total & $2{,}481$ & $100\%$ \\
\bottomrule[1.25pt]
\end{tabular*}
\end{center}
\end{table}

\section{Ethical Discussion}
\label{sec:ethic}

Audio-driven facial motion synthesis offers clear benefits in video conferencing, accessibility tools for camera-off communication, virtual production, and the entertainment industry, but the same capability can enable identity impersonation, harassment, and the spread of disinformation if used irresponsibly. \proposed amplifies these risks in two ways: it requires only an audio stream plus a handful of casually recorded reference clips, lowering the barrier relative to image-driven deepfakes; and its zero-lookahead pipeline can in principle support synchronous impersonation in live video calls when paired with a voice-cloning system, a scenario that previously required heavy buffering. To mitigate misuse, our release will pair the code, model weights, and any synthetic data with three layers of safeguards. We will include a visible synthetic-content badge on every released demonstration video.
Our trained checkpoints will be released under a license that prohibits the commercial synthesis of identifiable individuals without their explicit consent and requires reference clips to be of the user themselves or of subjects who have given clear permission. Finally, we will share our synthetic data with the deepfake-detection community, with the two evaluation networks described in \S\ref{sec:supp_evals} serving as reference baselines for sync-based and identity-similarity detection.

\section{Limitations and Future Work}
\label{sec:limi}

Although \proposed delivers strong lip-sync, identity fidelity, and perceived realism under strict zero-lookahead constraints, several limitations remain. Under zero audio lookahead, the model has no advance signal that the user is about to speak, so the first one or two frames after a long silence lag behind a non-causal baseline; a small lookahead window of a few frames recovers most of the lost accuracy at the cost of tens of milliseconds of buffering. The expressiveness of the retrieved style prior is bounded by the diversity of the user's reference clips, and when the library covers only a narrow band of context the output tends to revert to the speaker's most frequent style; performance plateaus quickly with library size, but bridging the gap to truly out-of-distribution emotional expressions remains an open problem. Even with rich templates, audio alone cannot resolve every idiosyncratic facial behavior, particularly purely visual cues such as eye-contact direction or context-dependent reactions. Future work will explore generative motion priors to reduce template dependence and provide a graceful fallback for small libraries, a streaming extension to full-body motion for holistic telepresence avatars, a multilingual and paralinguistic extension covering events such as laughter and back-channel cues.

\paragraph{Input robustness.}
Input robustness involves both the driving audio and the reference videos. For driving audio, TalkVid contains in-the-wild recordings with ambient noise, including street interviews, exposing the model during training to noisy speech rather than only studio-clean recordings and providing robustness to common background noise. For reference videos, robustness to clutter and occlusion is primarily determined by the upstream motion tracker. Our method operates directly on the extracted motion vectors; existing trackers generally handle cluttered backgrounds and mild occlusions, while severe tracking errors may propagate into the reference library.

\end{document}